\def\cm{{\rm\thinspace cm}}

\def\erg{{\rm\thinspace erg}}

\def\K{{\rm\thinspace K}}
\def\keV{{\rm\thinspace keV}}
\def\km{{\rm\thinspace km}}
\def\kpc{{\rm\thinspace kpc}}

\def\Mpc{{\rm\thinspace Mpc}}
\def\Msun{\hbox{$\rm\thinspace M_{\odot}$}}

\def\s{{\rm\thinspace s}}

\def\cmps{\hbox{$\cm\s^{-1}\,$}}

\def\pcmcu{\hbox{$\cm^{-3}\,$}}

\def\ergpcmsqps{\hbox{$\erg\cm^{-2}\s^{-1}\,$}}

\def\ergps{\hbox{$\erg\s^{-1}\,$}}

\def\kmps{\hbox{$\km\s^{-1}\,$}}

\def\pcmsq{\hbox{$\cm^{-2}\,$}}

\def\ps{\hbox{$\s^{-1}\,$}}

\def\mdot{\hbox{$\dot{m}$}}
\def\Mdot{\hbox{$\dot{M}$}}
\def\spose#1{\hbox to 0pt{#1\hss}}
\def\approxlt{\mathrel{\spose{\lower 3pt\hbox{$\sim$}}
        \raise 2.0pt\hbox{$<$}}}
\def\approxgt{\mathrel{\spose{\lower 3pt\hbox{$\sim$}}
        \raise 2.0pt\hbox{$>$}}}

\documentstyle[psfig]{mn}

\title[ADAFs in elliptical galaxy cores]
{Strong observational constraints on Advection-Dominated 
Accretion in the cores of elliptical galaxies}

\author[T.~Di~Matteo et al.]
{T.~Di~Matteo$^1$, A.~C.~Fabian$^1$, M.~J.~Rees$^1$, C.~L.~Carilli$^2$ and R.~J.~Ivison$^3$\\ 
{\small $^1$Institute of Astronomy, Madingley Road,
Cambridge, CB3 OHA }\\
{\small $^2$ NRAO, P.O. Box O, Socorro, NM, 87801, USA}\\
{\small $^3$ Institute for Astronomy, Dept. of Physics \& Astronomy, University
of Edinburgh, Blackford Hill, Edinburgh EH9 3HJ}}

\begin{document}

\maketitle

\begin{abstract}
The growing evidence for supermassive black holes in the centres of
relatively nearby galaxies has brought into sharper focus the question
of why elliptical galaxies, rich in hot gas, do not possess
quasar--like luminosities.  Recent studies suggest that the presence
of advection--dominated accretion flows (ADAFs) with their associated
low radiative efficiency, might provide a very promising explanation
for the observed quiescence of these systems.  Although ADAF models
have been applied to a number of low--luminosity systems compelling
observational evidence for their existence is still required.  Here,
we examine new high-frequency radio observations of the three giant,
low-luminosity elliptical galaxies NGC 4649, NGC 4472 and NGC 4636
obtained using the Very Large Array (VLA) and the sub-millimetre
common-user bolometer array (SCUBA) on the James Clerk Maxwell
Telescope (JCMT). At these frequencies the predictions are very
precise and an ADAF is unequivocally characterised by a slowly rising
spectrum with a sharp spectral cut--off produced by thermal
synchrotron radiation. Although X-ray analysis of these galaxy cores
provides very strong clues for their extreme quiescence (and makes the
case of advective--accretion plausible) the new radio limits disagree
severely with the canonical ADAF predictions which significantly
overestimate the observed flux. While the present observations do not
rule out the presence of an ADAF in the systems considered here, they
do place strong constraints on the model. If the accretion in these
objects occurs in an advection--dominated mode then our radio limits
imply that the emission from their central regions must be
suppressed. We examine the possibility that the magnetic field in the flow
is extremely low or that synchrotron emission is free--free absorbed
by cold material in the accretion flow. We also discuss whether slow
non--radiating accretion flows may drive winds/outflows to remove
energy, angular momentum and mass so that the central densities,
pressure and emissivities are much smaller than in a standard ADAF.
\end{abstract}


\section{Introduction}

Clear evidence for supermassive black holes comes from studies of
nearby galaxies.  The centres of most of these galaxies display little
or no activity, but most seem to harbour massive
black holes probably remnants of an earlier quasar phase (Kormendy
and Richstone 1995, Tremaine 1997 for recent reviews). It is a puzzle
why the massive black holes in the nearby galaxies do not show
quasar--like activity. The problem becomes particularly relevant for the
case of nearby giant elliptical galaxies where it is implausible to
postulate that the black holes are 'starved' of fuel to power the
accretion process.  X-ray studies for such giant ellipticals reveal
the presence of an extensive hot gas pervading their centres (as well
as the surrounding cluster). If there were a huge black hole then some
of this gas would inevitably be accreting into it, at a rate which can
be estimated from the Bondi (1952) formula (a lower limit). This
accretion would give rise to far more activity than is observed if
the radiative efficiency were as high as 10 per cent as generally
postulated in standard accretion theory (Fabian \& Canizares 1988).

Motivated by this problem Fabian \& Rees (1995) have recently
suggested that the final stages of accretion in such objects occurs
through an advection--dominated accretion flow (ADAF; Narayan \& Yi
1995, Abramovicz et al. 1995). Within the context of such an accretion
mode, the quiescence of these nuclei is not surprising. For low
accretion rates, the expected luminosity scales as $\dot{M}^2$ rather
than $\dot{M}$: when the accretion occurs at a low rate, and the
viscosity is high enough to ensure that the gas accretes quickly (and
densities are low), the radiative efficiency is low. The gas
inflates into a thick torus where the kinetic temperature of the ions
is close to the virial temperature. Only a small fraction of the
gravitational energy is radiated; most is swallowed into the black
hole.

The relevance of this accretion solution in ellipticals has been
illustrated by its application to the giant elliptical galaxy M87 (NGC
4486).  This has a quiescent active nucleus the luminosity of which
does not exceed $10^{42} \ergps$ with a nuclear black hole mass
$M\approx 3\times 10^9
\Msun$ (Ford et al. 1994; Harms et al. 1994, Macchetto et al. 1997). 
The radio and X-ray emission in M87 is fully consistent with accretion
at the expected rate only if it involves an ADAF (Reynolds et
al. 1996).  Furthermore, it has been shown that if elliptical galaxies
like M60 (NGG 4649, Di Matteo \& Fabian 1997) or NGC 4472, NGC 4636
(Mahadevan 1997) have advection--dominated nuclei then the upper limit
on their nuclear mass black hole can be of the order of $M \sim 10^9
\Msun$, thereby allowing supermassive black holes in these systems.

Although accretion of hot gas in an elliptical galaxy may create the
ideal circumstances for an ADAF to operate, stronger observational
evidence needs to be obtained. In this paper we examine how ADAF
models of nuclei in ellipticals can be constrained with further high
frequency radio and submillimeter data.

\subsection{Testing the ADAF paradigm for elliptical galaxies}
In an ADAF, the majority of the observable emission is in the radio
and X--ray bands.  In the radio band the emission results from
cyclo--synchrotron emission due to the strong magnetic field in the
inner parts of the accretion flow. The X-ray emission is due either to
bremsstrahlung or inverse Compton scattering.  On the observational
side, therefore, the cleanest way of determining the presence of an
ADAF is to examine the correlation between radio and X-ray
emissions. The spectral predictions of the ADAF model are quite
precise in these wavebands and the expected correlation tight. Weak
central radio sources are observed in the centres of many otherwise
quiescent ellipticals (Sadler et al. 1989). If these galaxies all
harbour massive black holes, this radio emission could similarly be
attributed to the synchrotron radiation in an inefficient accretion mode
(i.e. an ADAF). 

The self--absorbed synchrotron spectrum in an ADAF slowly rises with
frequency in the radio band, up to some critical turnover frequency,
typically in high radio to sub--mm frequencies, above which it should
abruptly drop. A large spectral break there is crucially indicative of
thermal self--absorbed synchrotron emission.  Observations of a
slowly--rising spectrum which sharply cuts off in the sub-mm band,
would provide the cleanest evidence that such an accretion mode is
taking place in nearby galaxies.

Here, we examine the high-frequency Very Large Array (VLA) and
Submillimeter Common User Bolometer Array (SCUBA) data for three giant
ellipticals in the Virgo cluster NGC 4649, NGC 4636, and NGC 4472
where the spectral turnover should be seen.  Slee et al. (1994) show
that the size of most of the radio cores in elliptical galaxies is on
the parsec scale or less which would imply that they are powered by an
engine similar to that in more luminous radio galaxies.  It is
interesting to note that these authors find that the cores of
ellipticals usually have flat or rising spectra. The motivation for
the choice of these galaxies is two--fold.  On one hand, the 
radio data already existing (Fabbiano et al. 1988) for both
NGC 4649 and NGC 4636 imply a flat/slowly rising radio spectral
component and therefore a tentative agreement with the ADAF
predictions in the intermediate--frequency radio band.  In particular
the radio spectrum of NGC 4649 was shown to closely resemble that
expected from the self--absorbed synchrotron radiation (Di Matteo \&
Fabian 1997).

On the other hand, as discussed above, these giant elliptical galaxies
seem to be the strongest candidates for illustrating the possibility
of explaining quiescent accretion via ADAFs.  Recent {\it Hubble Space
Telescope (HST)} observations (Magorrian et al. 1998) have now
provided high resolution dynamical models for these and many other
nearby galaxies and a determination of nuclear black hole masses 
of the order of $10^8-10^9 \Msun$ (see Table
~\ref{t:bh}).

\begin{table}
\caption{Black hole masses (Magorrian et al. 1998) and predicted luminosities}
\label{t:bh}
\begin{center}
\begin{tabular}{ccc}\hline
Object & Black Hole & $ L_{\rm predicted}$ \\
\vspace{0.3cm}& ($10^9 \Msun$) & ($10^{44} \ergps$) \\
\hline
NGC 4649 & 3.9 & 5.1 \\

NGC 4472 & 2.6 &  2.3 \\

NGC 4636 & 0.3 &  0.03\\

\hline
\end{tabular}
\end{center}
\end{table}
A deprojection analysis of data from the {\it ROSAT} High Resolution
Imager (HRI) shows that the hot interstellar medium (ISM) in NGC 4649,
4472 and 4636 has a central density $n\approx 0.1\pcmcu$ for a sound
speed $c_{\rm s} \sim 300\kmps$.  The resulting Bondi accretion rate
onto a central black hole of $10^9 \Msun$ is then
\begin{equation}
\dot{M}_{Bondi}\sim 10^{-2} \left(\frac{M}{10^9\Msun}\right)^2\,\left(\frac{n_{\infty}}
{0.1}\right)\left(\frac{300}{c_{\infty}}\right)^3 \;\;\Msun {\rm yr^{-1}}.
\end{equation}
For the black hole masses predicted for these galaxies (Table
~\ref{t:bh}; Magorrian et al. 1998) the Bondi rate, for a radiative
efficiency of $\eta=0.1$ (as assumed for standard accretion) predicts
a luminosity of the order of $L_{\rm predicted}$ exceeding
$10^{44}\ergps$ for two of them (see Table
~\ref{t:bh}). Observationally, the nuclei of these giant ellipticals
are orders of magnitude less active (see Fig. ~\ref{4649spec} and
~\ref{4472_4636specs}). The observed luminosity of their cores does not
exceed $10^{40}\ergps$ over all energies.
\begin{figure*}
\centerline{
\hbox{
\psfig{figure=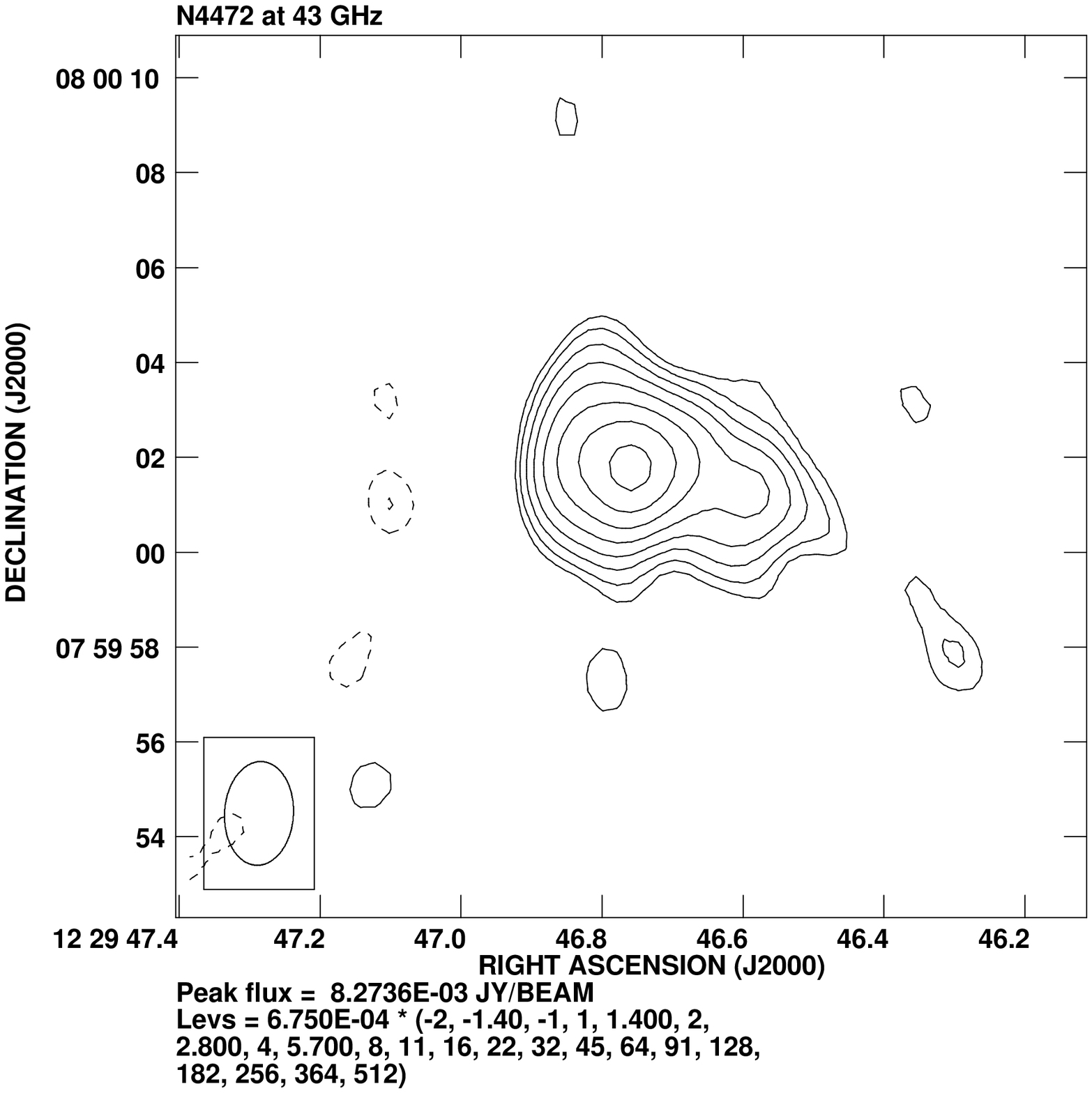,width=0.35\textwidth}
\psfig{figure=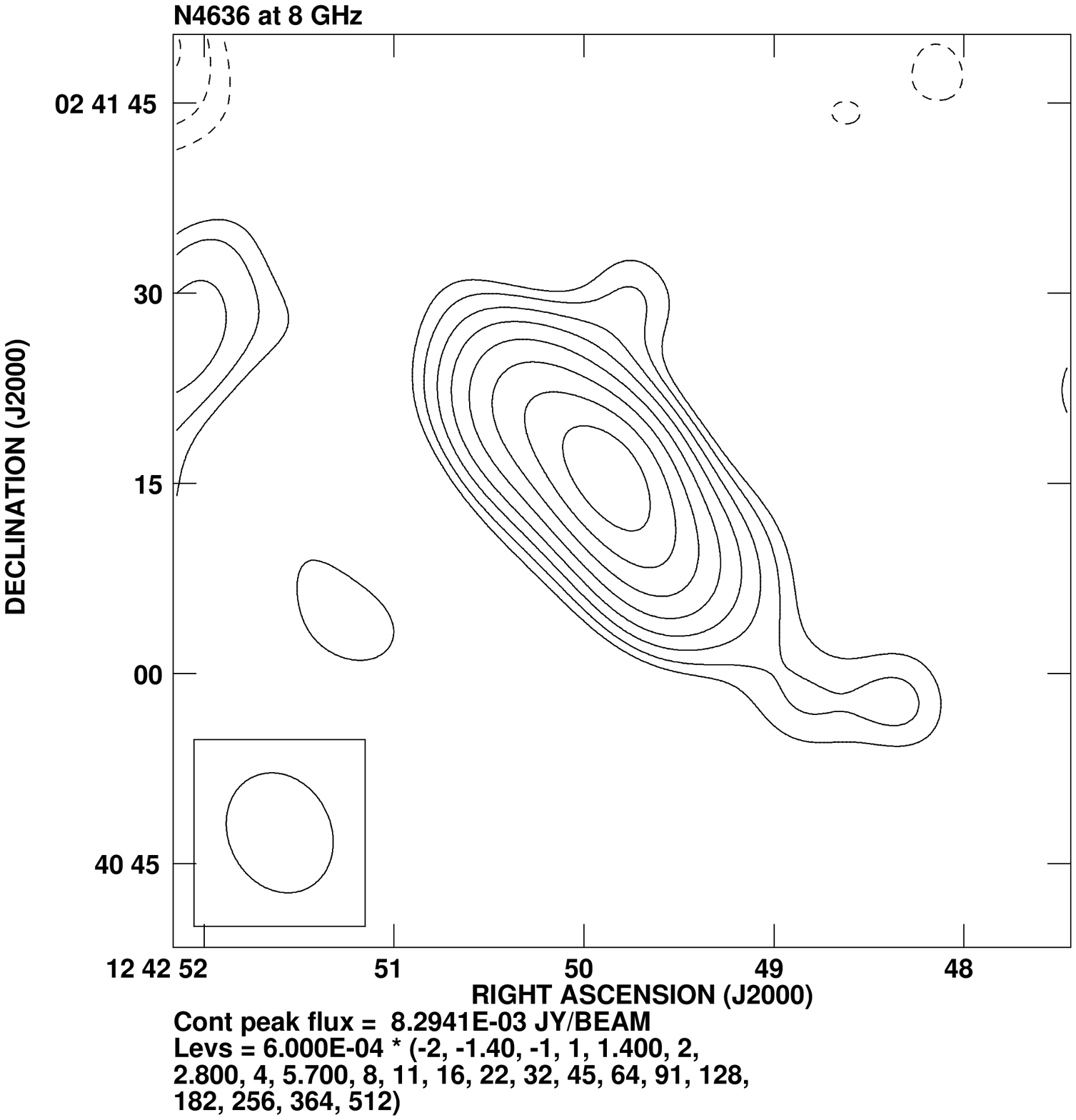,width=0.335\textwidth}
\psfig{figure=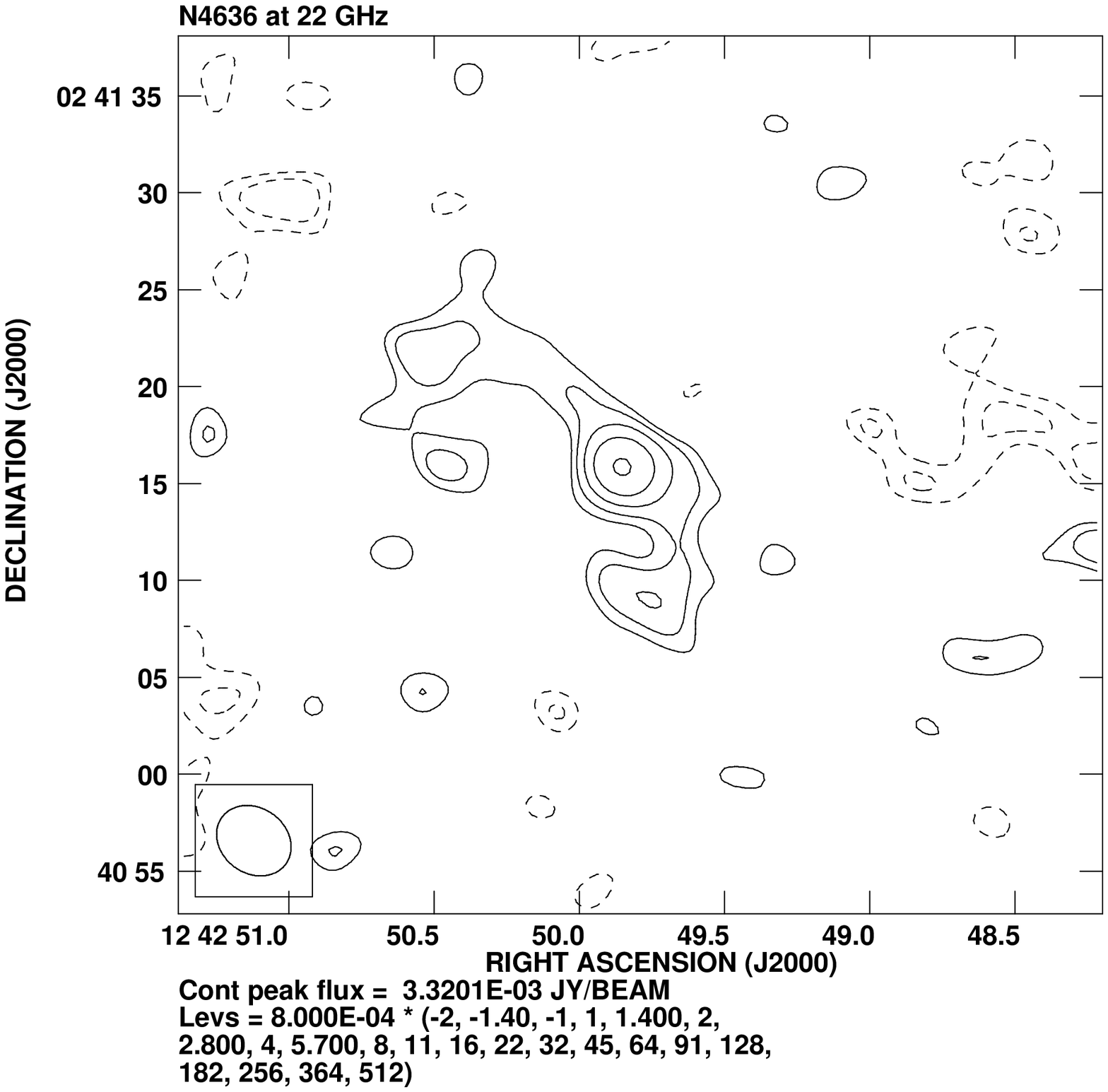,width=0.35\textwidth}
\newline
}}
\caption{VLA high resolution radio images for NGC 4472 at 43 GHz and NGC 4636 at 8 and 22 GHz. The FWHM of the restoring Gaussian beam 
is shown in the box in lower left corner. These are the two extended
sources in the sample.}
\label{radiomaps}
\end{figure*}
In this paper we test the ADAF paradigm for elliptical
galaxies by analysing high frequency radio and sub--mm data together with X-ray
{\it ROSAT} HRI observations.  In Section 2, we show our VLA and SCUBA
data, present upper limits for the X-ray flux from the core, and
summarize some other relevant data from the literature on the
full--band spectrum of the cores of NGC 4649, 4636 and 4472.  Section 3
describes some details of our ADAF model spectrum calculation. In
Section 4, we compare the model spectrum with the data and find
very strong new constraints for the ADAF models. Section 5, discusses
some important implications of our results.

As mentioned above, demonstrating that
accretion via ADAFs is operating in the nearby universe would be an
important step in understanding the demise of quasars.

\begin{table}
\caption{SCUBA observations.}
\label{t:scuba}
\begin{center}
\begin{tabular}{ccccc}\hline
Object & Wavelength &  Flux  \\
\vspace{0.3cm}& ${\rm mm}$ & {\rm mJy} & \vspace{0.3cm} \\ 
\hline
NGC 4649  &2.0&	  $1.8 \pm 1.7$  \\     

NGC 4472 & 2.0 &  $9.6 \pm 2.4$  \\
	 & 1.35 &  $4.5 \pm 1.1$  \\ 
	 & 0.85 &  $3.0 \pm 1.3$ \\

NGC 4636 & --  \\

\hline
\end{tabular}
\end{center}
\end{table}

\begin{table*}
\caption{VLA data}
\label{t:vla}
\begin{center}
\begin{tabular}{ccccccc}\hline
Galaxy & Frequency & RMS & Total & Peak & Peak& notes\\
\vspace{0.3cm} &&&&&at 8GHz resolution& \\
\vspace{0.3cm}& $\nu$ (GHz)& {\rm mJy}& $F_{\nu}$({\rm mJy}) &$F_{\nu}$({\rm mJy}) & $F_{\nu}$({\rm mJy}) &   \vspace{0.3cm} \\ 
\hline
NGC 4649 &  8.4& 0.10 & $18.1\pm0.9$ & $17.6\pm0.9$ &  --       & point source\\
	 &   22& 0.30 & $23.5\pm1.4$& $ 21.9\pm1.3$  &  --      & \\
	 &  43 & 0.27 & $12.9\pm1.3 $& $12.7\pm1.3$ &  --      & \\   
NGC 4472 &  8.4& 0.30 & $49.0\pm2.5$ & $44.0\pm 2.2 $& $44.0\pm2.2$ & extended source \\
	 &  22 & 0.32 & $34.0\pm2.0$ &$ 23.9\pm1.4 $ &$ 31.9\pm1.9$ & \\
	 &  43 & 0.22 &$ 23.5\pm2.4 $& $ 8.3\pm0.8$  &$ 20.7\pm1.7$ & \\
NGC 4636 & 8.4 &0.33  & $18.1\pm0.9$ &  $8.3\pm0.4$  & $8.3\pm0.4$  & extended source \\
	 & 22  &0.48  & $13.5\pm1.4$ & $ 3.3\pm0.5$  & $5.6\pm1.2$  & \\ 
	 & 43  &0.22  &  --        &  $0.6\pm0.2$  & $<1.8 $      &\\
\hline
\end{tabular}
\end{center}
\end{table*}

\section{The spectrum of the core emission}
In order to examine the nature of the accretion flow in elliptical
galaxies, we have compiled the best observational limits on the full
band spectrum of the core emission.  Our aim is to obtain good
observational limits on the core flux over a wide range of frequencies
rather than to compile a comprehensive list of all previous
observations.  Some contribution from the weak jets and in particular
from the underlying galaxy are unavoidable and so the derived spectrum
should be considered an upper limit to that of the accretion flow at
the elliptical galaxies cores.

In the next two sections we present our new high
frequency radio and sub--mm (VLA and SCUBA respectively) observations
and ROSAT HRI upper limits to the X-ray emission.  High radio
frequency and sub--mm observations together with X-ray measurements
are crucial for constraining the case of ADAFs in elliptical galaxy
cores.  The data for NGC 4649, NGC 4472 and NGC 4636 are summarized in
Table ~\ref{t:4649},~\ref{t:4472} and ~\ref{t:4636} respectively.

\subsection{The radio data}
Radio continuum surveys of elliptical and S0 galaxies have shown that
the sources in radio--quiet galaxies tend to be compact with
relatively flat or slowly rising radio spectra (with typical spectral
index of 0.3--0.4), suggesting that the radio emission from
ellipticals are in general of nuclear origin (Slee et al. 1994; Wrobel
1991) and might be powered by engines similar to those in more
luminous radio galaxies.

Very Large Array (VLA) imaging by Stanger \& Warwick (1986) and Wrobel
\& Heeshen (1991; at $\nu = 5$ GHz) has shown 
that the nuclear radio sources in NGC 4649, NGC 4472 and NGC 4636 are
extended with a compact ($\approxlt$ 4 arc--sec), core component. The
extended structure defines radio lobes powered by weak jets with
linear dimensions $\sim 2 {\rm
\kpc}$.  
The relatively flat/slowly--rising spectral nature of the
high--frequency emission from NGC 4636 and NGC 4649, which is well
defined by the data of Fabbiano et al. (1987) (see Table ~\ref{t:4649}
and ~\ref{t:4636}), motivated the selection of these galaxies for
further higher frequency observations.  NGC 4472 also follows the same
general trend (Wrobel \& Heeshen 1991)
~~~~~~~~
\begin{figure}
 \centerline{\psfig{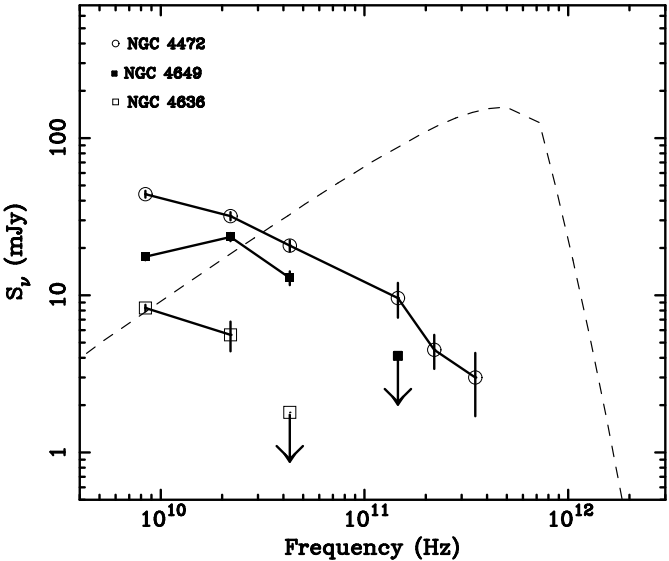}}
 \caption{The VLA (8.4, 22, 43 GHz) and SCUBA sub--mm flux
 measurements for NGC 4649, NGC 4472, and NGC 4636 (joined by a solid
 line). NGC 4472 is the brightest source in the sample, its spectrum
 is quite steep and probably dominated by jet emission up to sub--mm
 wavelengths (see also the 43 GHz image, Fig.~\ref{radiomaps}). The
 spectrum of NGC 4649 rises up to the 22 GHz but by 43 GHz the
 spectrum has turned over. SCUBA non--detection at $2000 \mu$ is
 represented as a 3$\sigma$ upper limit. No SCUBA data are available
 for NGC 4636 but VLA measurements imply a flat spectrum with
 (possibly) a sharp cut off ($3\sigma$ upper limit at 43 GHz). The
 synchrotron emission as predicted by the standard ADAF model
 (dashed--line; with typically $m=10^9$, $\mdot=1\times 10^{-3}$,
 $\beta=0.5$, $\alpha=0.3$) greatly overestimates the total observed
 flux. The model is inconsistent with the radio limits for all three of the
 sources.}
\label{vla_scuba_fig}
\end{figure}
~~~~~~

\begin{figure*}
\hbox{
\hspace{-0.1cm}\psfig{figure=psf_point.ps,width=0.5\textwidth,angle=270}
\hspace{0.2cm}\psfig{figure=psf_point_n4636.ps,width=0.5\textwidth,angle=270}

}
\vspace{0.5cm}
\hbox{
\hspace{-0.1cm}\psfig{figure=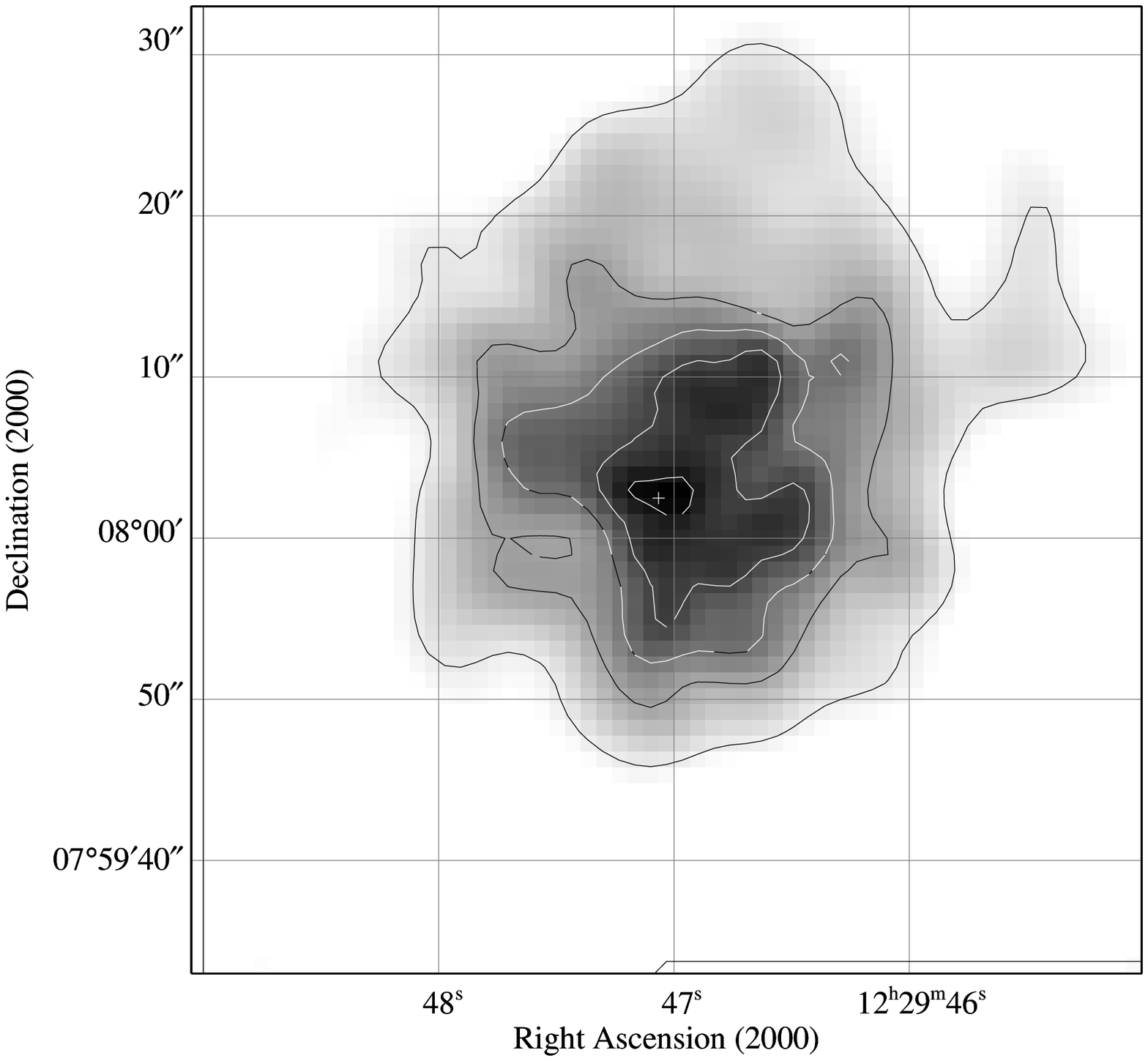,width=0.5\textwidth}
\hspace{0.2cm}\psfig{figure=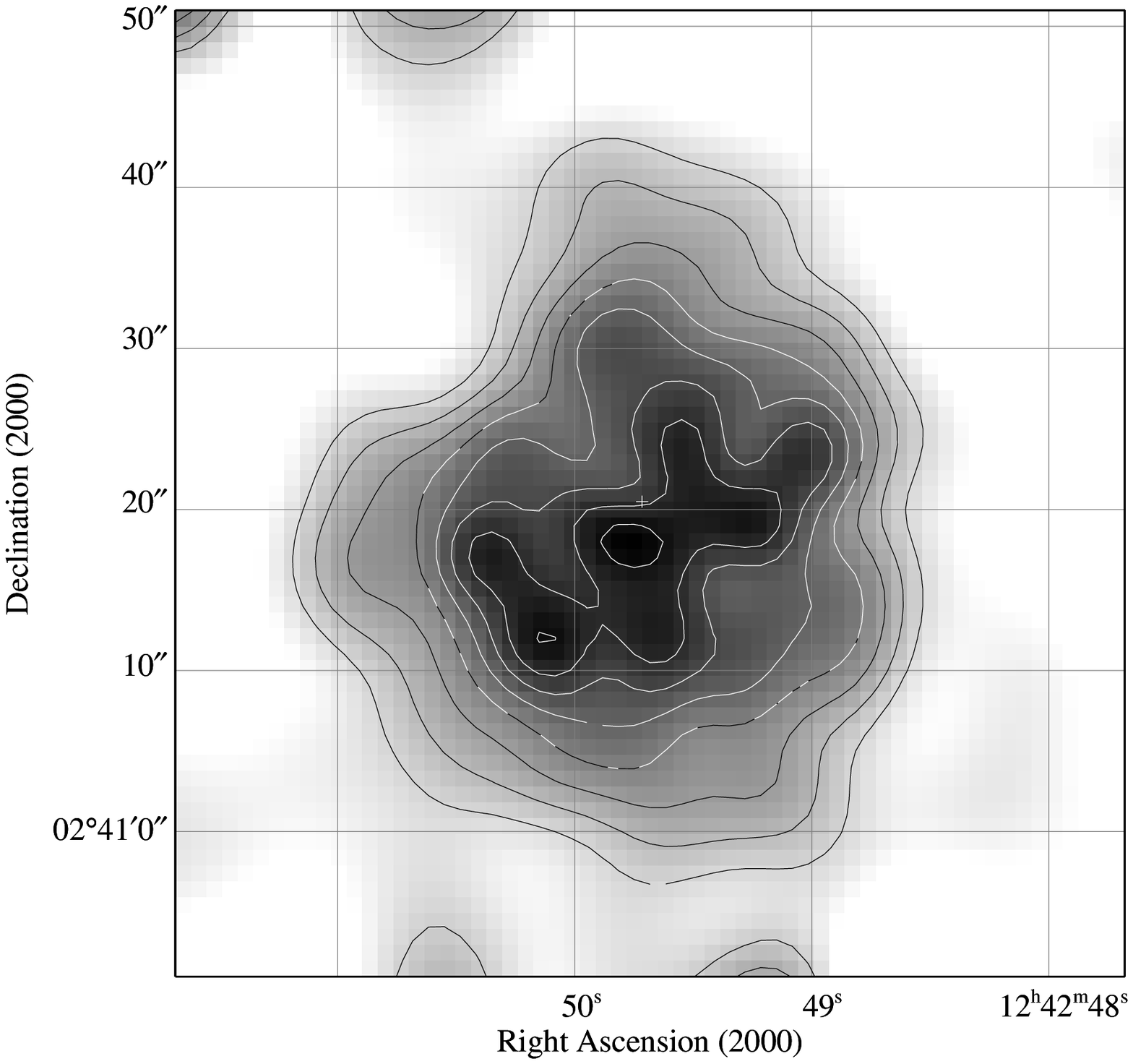,width=0.5\textwidth}
}

\caption{The core regions of NGC 4472 (on the right) and NGC 3636 
(on the left) as imaged in 27 ks and 22 ks exposures respectively with
the {\it ROSAT} HRI. The distribution of the X-ray emission of the
central region of both galaxies is quite complex and asymmetric.  The
diffuse emission is from the hot interstellar medium. The images have
been smoothed adaptively prior to contouring; the minimum number of
counts over which smoothing occurred is 20.  Contour levels are
equally spaced on a logarithmic scale, starting at $3\times10^{-3}
{\rm count\; s^{-1} arcmin^{-2}}$ and increasing by a factor 1.6
between adjacent contours (NGC 4472) and starting at $8\times10^{-4}
{\rm count\; s^{-1} arcmin^{-2}}$ and increasing by a factor 1.3 (NGC
4636).  On top, we also show the radial distribution of the surface
brightness and the contribution of a point source to their emission.
The solid line shows the PSF of the {\it ROSAT} HRI compared to the
total count rate per unit area.}
\label{xrayfigs}
\end{figure*}

\subsection{VLA and SCUBA Observations}
Measurements of all three galaxies were obtained with the Very Large
Array (VLA) at 8.4, 22 and 43 GHz during December 1997.  The resulting
data are shown in Table ~\ref{t:vla}. NGC 4472 and 4636 are extended
sources and their high resolution radio images are shown in
Fig. ~\ref{radiomaps}. In order to obtain the best limits on the core
flux for the extended sources, the flux densities were derived from
images convolved to the resolution of the 8 GHz image at all
frequencies (fifth column in Table ~\ref{t:vla}). The
full--width--at--half--maximum (FWHM) of Gaussian restoring beam at 8
GHz ($\delta = -8$deg) was 9x8 arcsec$^2$. The flux scale was set using
3C 286.

On January 30,
1998, NCG 4649 and NGC 4472 were also observed in the submillimeter
band using the new SCUBA bolometer camera (Holland et al.\ 1998;
Robson et al.\ 1998) on the JCMT. Data were obtained at 450, 850, 1350
and 2000\,$\mu$m during a good, stable night. The photometry scans
were interspersed with checks on the accuracy of the pointing and flux
calibration.  The SCUBA data were analysed using the dedicated SURF
reduction package (Jenness 1998) using the techniques described by
Ivison et al.\ (1998a, 1998b). The SCUBA flux measurements, are summarized
in Table ~\ref{t:scuba}. Both VLA and SCUBA data are plotted in
Fig. ~\ref{vla_scuba_fig}.

{\bf NGC 4649} -- In accordance with the previous observations (Fabbiano
et al. 1987; Wrobel et al. 1991) our VLA measurements imply that this
source is strongly core dominated. Its spectrum, as measured by VLA
and SCUBA shows evidence of a slowly rising, high frequency component
up to $\sim 30$ GHz.  The spectrum then sharply turns over, as implied
by the 43 Hz measurement and the lack of a SCUBA detection 
(represented as 3$\sigma$ upper limit) at 2000 $\mu$ m.

{\bf NGC 4472} -- The steeper spectral slope
(Fig.~\ref{vla_scuba_fig}) is probably due to the dominance of
emission from a jet--like structure. Even at 43 GHz
(Fig. ~\ref{radiomaps}) the emission is still fairly extended. The
extended component probably dominates all the way into the SCUBA
frequencies.

{\bf NGC 4636} -- This galaxy showed the presence of a flat component in
its spectrum in the 1.4 and 4.75 GHz measurements of Fabbiano et
al. (1987). The relative contributions of the core components in this
source is clearly illustrated in the maps of our 8 GHz and 22 GHz VLA
observations, shown in Fig. ~\ref{radiomaps}.  Our VLA high--frequency
observations imply a relatively sharp turn over in the spectrum at
around 10 GHz. At 43 GHz only a ($3\sigma$) upper limit could be set
to the the source flux. Sub--millimetre observations of this object
were not performed.

\subsection{The X-ray data}

{\it Einstein} Observatory observations of giant elliptical galaxies
reveal that the overall X-ray emission is generally extended with most
of it arising from the hot gas at ${\rm kT} \sim 1{\rm \keV}$
(Fabbiano, Kim \& Trinchieri, 1992).

Here we examine {\it ROSAT} HRI datasets in order to constrain the
nuclear X--ray flux of NGC 4472 and NGC 4636 (see Di Matteo \& Fabian,
1997 for the analysis of NGC 4649). Fig. ~\ref{xrayfigs} shows the HRI
image of NGC 4472 and NGC 4636.  Consistent with the {\it Einstein}
HRI observations, the {\it ROSAT} HRI image shows that these sources
are extended with most of the emission coming from hot gas in the
interstellar medium.  The low luminosity radio activity though,
indicates that there is some form of active central engine in the core
of these galaxies.  In order to find an upper limit to their nuclear
emission (i.e. a maximum possible luminosity of a point source) we fit
the surface brightness profile with an extended component modelled by
a King profile plus a point source modified by the Point Spread
Function (PSF) of the HRI (see Fig. ~\ref{xrayfigs}).

{\bf NGC 4472} -- We analyse the {\it ROSAT} HRI data
resulting from a 27\,367\,s exposure.  Our formal $3\sigma$ upper
limit, assuming no intrinsic absorption, predicts a count rate from the
point source component of $2.1\times 10^{-3}$\,ct\,s$^{-1}$ (Fig. ~\ref{xrayfigs}).
Assuming the spectrum to be a power-law with a canonical photon index
$\Gamma=2.0$ modified by the effects of Galactic absorption (with
column density $N_{\rm H}=3.0\times 10^{20}\pcmsq$); this count rate
implies a flux density at 1\,keV of $F(1\,{\rm
keV})=6.8\times10^{-14}\ergpcmsqps\keV^{-1}$.  This result is fairly
insensitive to the choice of power law index. 

{\bf NGC 4636} -- We analyze the joint {\it ROSAT} HRI data resulting
from 13\,264\,s and 8\,688\,s exposures.  The formal $3\sigma$ upper
limit assuming no intrinsic absorption predicts a count rate from the
point source component of $2.5\times 10^{-3}$\,ct\,s$^{-1}$
(Fig. \ref{xrayfigs}).  Assuming the same energy distribution
described above this count rate corresponds to a flux density at
1\,keV of $F(1\,{\rm keV})=7.8\times10^{-14}\ergpcmsqps\keV^{-1}$.

{\bf NGC 4649} --
see analogous analysis in Di Matteo \& Fabian (1997).

\section {ADAFs around supermassive black holes}
In an advection-dominated accretion flow (ADAF) most of the energy
released by viscous dissipation is stored within the gas and advected
inward with the accreted plasma; only a small fraction is radiated
away (see Narayan, Mahadevan \& Quataret 1998 for a recent
review). Work on ADAFs has been concerned with low-$\Mdot$
solutions (Rees 1982; Narayan \& Yi 1995a,b; Abramowicz et al. 1995)
which occur when the accretion rate is lower then a critical value,
$\mdot_{\rm crit} \approxlt 1.3
\alpha^2$, (hereafter $\dot{m}=\dot{M}/\dot{M}_{\rm
Edd}$ in Eddington units). This optically--thin branch of the solutions
makes use of the standard $\alpha$ viscosity and is based on certain
critical assumptions.  ADAF models assume that most of the viscous
energy is deposited into the ions and only a small fraction of energy
goes directly into the electrons.  The energy is then transferred from
the ions to the electrons via Coulomb collisions. Because Coulomb
transfer in the the low--density ADAF is inefficient, the plasma is
two-temperature. The kinetic temperature of the ions reaches the
virial temperature while the electron temperature saturates at around
$10^9-10^{10}\K$.

The spectrum from an ADAF is very different from that of a standard
thin disk.  The high electron and ion temperatures, the presence of an
equipartition magnetic field and the fact that the gas is optically
thin, allow a variety of radiation processes to contribute to the
emitted spectrum from radio to gamma rays.  The radio to X-ray
emission is produced by the electrons via synchrotron, bremsstrahlung
and inverse Compton processes. The gamma--ray radiation results from
the decay of neutral pions created in proton--proton collisions.  Note
that the radio emission from an ADAF is much stronger than that from a
standard thin disk.

In the remainder of this section we summarize how we compute the
spectrum of an ADAF based on the model of Narayan \& Yi 1995 and recent
calculations by Narayan, Barret \& McClintock 1997 (hereafter NBM) and
Esin, McClintock \& Narayan 1998, (hereafter ESM). 

\begin{table*}
\caption{Summary of data for the core of NGC 4649.}
\label{t:4649}
\begin{center}
\begin{tabular}{ccccc}\hline
Frequency & $\nu F_{\nu}$ & reference & notes \\
$\nu$ (Hz) & (10$^{-15}$\,erg\,s$^{-1}$\,cm$^{-2}$) & \vspace{0.3cm} \\
\hline 
$1.4\times 10^9$ & 0.37 & Hummel et al. (1983) & Westerbork \\
$4.75\times 10^9$ & 1.14 & Fabbiano  et al. (1987), Wrobel (1991) & Effelsberg, VLA\\
$8.4\times 10^9$  & 1.49 & this work & VLA\\
$1.07\times 10^{10}$ & 2.78 & Fabbiano et al. (1987) & Effelsberg \\
$2.2\times 10^{10}$ & 4.83 & this work & VLA \\
$3.3\times 10^{10}$  & 10.5 & Fabbiano et al. (1987) & Effelsberg \\
$4.3\times 10^{10}$ & 5.47 & this work &VLA \\
$1.5\times 10^{11}$ & $\le$ 7 & this work & SCUBA \\ 
$5.45\times 10^{14}$ & $\le$ 180 & Byun et al. (1996) & HST \\
$2.4\times 10^{17}$  & $\le$ 150 & Paper 1 & {\it ROSAT} HRI\\

\hline

\end{tabular}
\end{center}
\end{table*}
\begin{figure*}
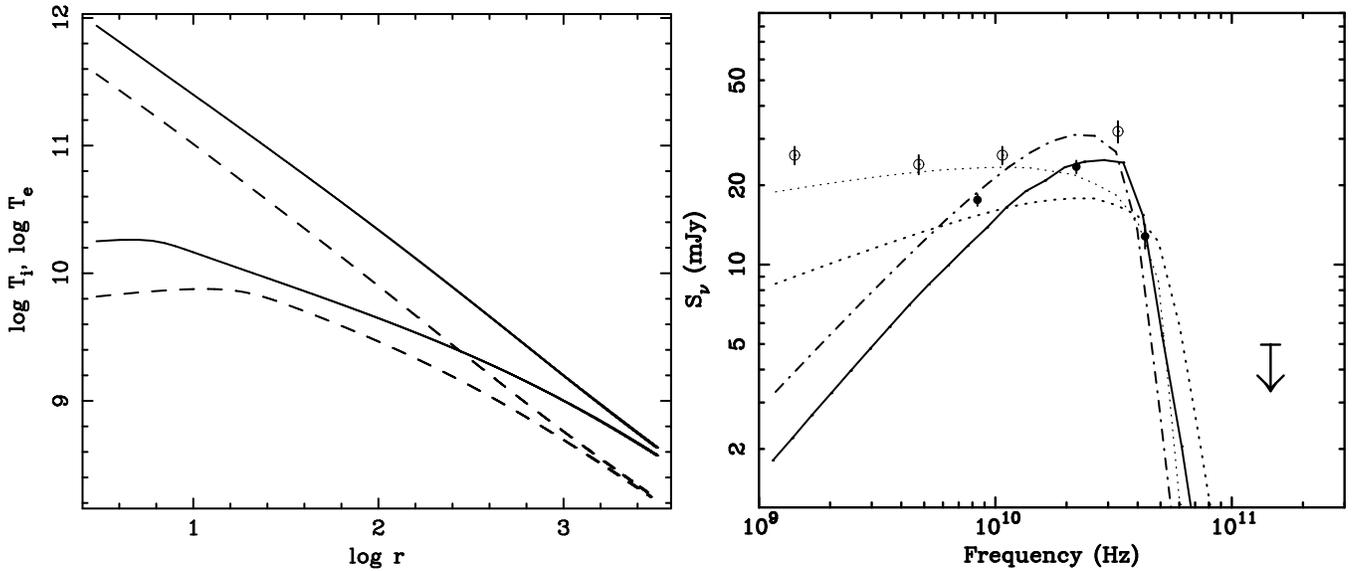

\centerline{
\hbox{
\psfig{figure=temp_prof.ps,width=0.5\textwidth,angle=270}
\psfig{figure=fig2.ps,width=0.5\textwidth,angle=270}
\newline
}}
\caption{On the left, the temperature profile
for an ADAF. The solid line is for $\beta \sim 0.99$ and the dashed
 line for $\beta=0.5$. This corresponds to the mass and accretion rate
 of NCG 4649: for the other objects the temperature solution is very
 similar to the one shown.  On the right, comparison between the
 synchrotron flux and the high energy radio and submillimeter limits
 for NGC 4649. The open circles represent the Fabbiano et al. (1987)
 measurements the filled circles and filled squares our VLA and SCUBA
 new limits. Our high energy measurements imply that the spectral
 turnover due to the self--absorbed synchrotron emission occurs at
 much lower frequencies than expected from previous a analysis (Di
 Matteo \& Fabian 1997) and in general form the canonical ADAF
 model. The model is consistent with the data only if $\beta \approxlt
 0.999$, i.e. a very low magnetic field (solid line), or if the
 emission is free--free absorbed by cold and dense material within
 $r_{\rm in} =17$. The dotted lines represent the case in which a
 magnetic wind carries away mass and angular momentum at large
 distances with $\mdot \propto (r/r_{\rm max})^{p}$.  The thicker
 dotted--line is for $r_{\rm max}=300$ and $p=1$, the dashed line for
 $r_{\rm max}=200$ and $p=1.2$.}
\label{4649temp_specfig}
\end{figure*}

\subsection{The model}
The modelling techniques have advanced significantly during the last
two years (NBM, ESM).  Here we briefly describe how the model of Di
Matteo \& Fabian (1997) has been improved by taking into account the
advection of energy by electrons which, as pointed out by Nakamura et
al. (1996) and Narayan, Barret \& McClintock (1997), can be important.
For convenience, we rescale the radial co-ordinate and define $r$ by
$r=R/R_{\rm S}$, where $R$ is the radial coordinate and $R_{\rm S}$ is
the Schwarzschild radius of the hole.

In order to describe the dynamics we use the local self--similar
solution calculated by Narayan \& Yi (1995); this provides a reasonably
accurate analytical estimate of the properties of the accretion flow.
The dynamical model is uniquely specified by four structure
parameters: $\alpha$, the viscosity parameter, $\beta$, the ratio of
magnetic to gas pressure, $f$, the advection parameter, which gives
the ratio of advected energy to viscous heat input and $\gamma$, the
adiabatic index of the fluid which is assumed to be a mixture of gas
and magnetic fields (Esin 1997); The models considered here are
extremely advection--dominated so that $f$ is virtually equal to 1 in
all cases.  For the present discussion the quantities of interest are:
\begin{eqnarray}
\!\!v\!\!\!&=&\! \! \!-2.1 \times 10^{10} \alpha c_{1} r^{-1/2} \cmps, 
\nonumber \\
n_e\!\!\!&=&\!\!\!2.0 \times 10^{19} c_{1}^{-1} c_3^{-1}
\alpha^{-1}m^{-1}\dot{m}r^{-3/2}\pcmcu, \nonumber \\
\!\!B\!\!\!&=&\!\!\!\!6.6 \times 10^8 \alpha^{-1/2}(1-\beta)^{1/2} 
m^{-1/2}\dot{m}^{1/2}r^{-5/4}c_1^{-1/2} c_3^{1/4} G, \nonumber \\ 
\tau_{\rm es}\!\!\!& =&\!\!\!12.4 \alpha^{-1} \dot{m}r^{-1/2} c_1^{-1}, 
\end{eqnarray}
where $v$ is the radial velocity, $n_{\rm e}$ is the number density of
electrons, $B$ is the magnetic field strength, $\tau_{\rm es}$ is the
the electron scattering optical depth, $m$ is the black hole mass in
solar units.  $c_1$ and $c_3$ are the constants defined in
Narayan \& Yi 1995 and are functions of $\gamma$ and $f$.

The thermodynamic state of the flow is described by the ion
temperature $T_{\rm i} \sim 10^{12} \beta r^{-1} \K$, which is to a very
good approximation the virial temperature, the electron temperature
$T_{\rm e}$ and the magnetic pressure, $p_{\rm mag} = (1 -\beta)p_{\rm
tot}$. It is usually assumed that all of the structural parameters in
the model are constant.  In order to determine the spectral properties
of the flow we need to calculate the electron temperature.  The
electron temperature profile is given by solving the electron energy balance
equation (Nakamura et al. 1996, NBM, EMB)
\begin{equation}
\rho T_{\rm e} v \frac{ds_{\rm e}}{dR}=\rho v \frac{du_{\rm e}}
{dR} -kT_{\rm e}v\frac{dn_{\rm e}}{dR}=Q^{\rm ie} + \delta Q^{+} - Q^{-},
\end{equation}
where $s_e$ is the entropy and the term on the left-hand side
represents the rate at which energy in the electrons is advected
inward in the flow (as shown by the two central terms, that is the
rate of change of internal energy $u_{\rm e}$ and of compression
heating/cooling). The other terms in equation Eqn. 3 are the rate at
which electrons are heated by Coulomb collisions, $Q^{\rm ie}$, the
total rate of viscous dissipation, $Q^{+}$ (we assume $\delta\sim
10^{-3}$), and the radiative cooling $Q^{-}$. This latter term
consists of electron--electron and ion--electron bremsstrahlung,
synchrotron emission and its Comptonization (as in Di Matteo \& Fabian
1997a).  With the simple scaling laws in Eqn. 3 the above equation is
a first order differential equation. We solve numerically the
temperature balance equation with one outer boundary condition
i.e. $Q^{\rm ie} = Q^{-}$ at $r=r_{\rm out} = 10^4$ (in the outer
regions the flow collapses to one temperature and radiative cooling --
dominated by bremsstrahlung in this regions -- balances the Coulomb
heating). Our approach for calculating the temperature profile is much
simpler than the procedure described NBM, EMN but it is in fairly good
agreement with their results.

Once the electron temperature profile has been determined (see
e.g. Fig.~\ref{4649temp_specfig}) we calculate the radiation spectrum of
an ADAF integrated over radius from $r_{\rm in}\sim 3$ to $r_{\rm out}
\sim 10^3$. This includes: synchrotron, bremsstrahlung and
Comptonization of the soft synchrotron photons. (Here the complicated
problem of the global radiation energy transfer, see e.g. NBM, ENM, is
not treated and local approximations are used).  As we discuss later,
the only tight constraints to the problem come from the radio data,
where an ADAF radiates via synchrotron emission.  In the thermal
plasma of an ADAF, synchrotron emission rises steeply with decreasing
frequency. Under most circumstances the emission becomes
self--absorbed and gives rise to a black--body spectrum below a
critical frequency $\nu_{\rm c}$. Above this frequency it decays
exponentially as expected from a thermal plasma, due to the
superposition of cyclotron harmonics. The synchrotron spectrum at each
radius is calculated in the way described by Di Matteo, Celotti \&
Fabian (1997), following the optical--thin formalism developed by
Mahadevan et al. (1996).  With the relevant information about the
original unscattered soft photon spectrum we can calculate the Compton
scattered spectrum using the scattering kernel and redistribution
functions derived by Poutanen (1994, and references therein). The
bremsstrahlung spectrum in this semi--relativistic regime is best
estimated with Stepney \& Guilbert (1983) formalism.  We also compute
pion production by the hot protons and the resulting $\gamma$--ray
emission through pion decay (see Appendix A and also Mahadevan et
al. 1997).

\section{Results}
In this section we combine the modelling techniques outlined in the
Section 3 and the data, with particular regards to the radio limits,
from Section 2 to constrain the ADAF models in the context of
elliptical galaxies.  

\begin{figure}
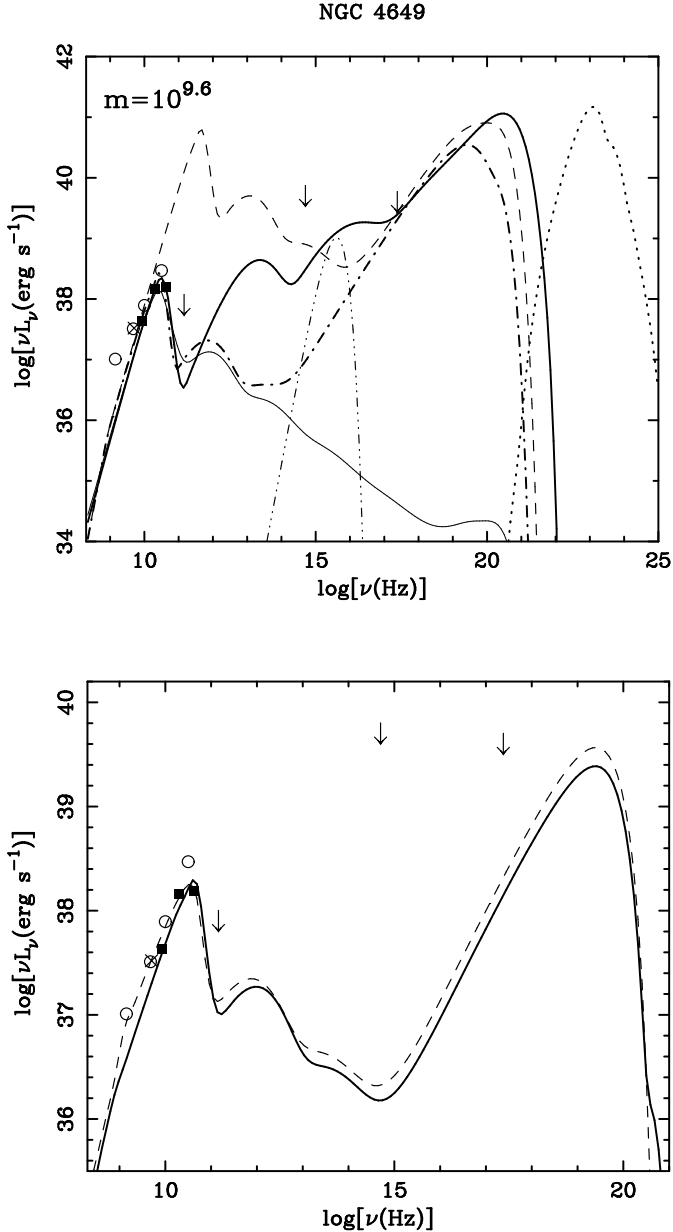

 \centerline{
\vbox{
\psfig{figure=fig3.ps,width=0.5\textwidth,angle=270}
\vspace{1cm}
\psfig{figure=fitall1.ps,width=0.5\textwidth,angle=270}
}}
\caption{On top, 
the full band spectra calculated with an ADAF model. The same mass,
 $m=10^{9.6}$, viscosity parameter, $\alpha=0.3$ and respective Bondi
 rates are assumed for all of the different models (apart from
 (iv)). Four models are shown: (i) one for equipartion magnetic field,
 $\beta=0.5$ --dashed line, clearly inconsistent with the radio limits
 (ii) one for $\beta =0.999$ solid line, (iii) one for which the radio
 emission is free--free absorbed within $r_{\rm in}=17$, dashed--dot
 line and dash--dot--dot line (representative blackbody emission from
 the cold absorbing material), (iv) and one for which $\alpha=0.3$,
 $\beta=0.5$ and $\mdot \sim 10^{-6}$, thin solid line.  At the
 bottom, the magnetic wind model with $\mdot \propto (r/r_{\rm
 max})^{p}$.  The solid line is for $r_{\rm max}=300$ and $p=1$, the dashed
 line for $r_{\rm max}=200$ and $p=1.2$. A distance of $15.8\Mpc$ is
 assumed. X-ray cooling processes dominate the emission.}

\label{4649spec}
\end{figure}

\subsection{The inconsistency between the radio data and the 'canonical' 
model predictions} 
Before presenting detailed models for the three
sources and discussing each one of them separately, we would like to 
illustrate (more qualitatively) the apparent inconsistency between the
ADAF predictions and our radio measurements. In Fig. ~\ref{vla_scuba_fig}
we show our VLA and SCUBA data points and the expected ADAF
self--absorbed synchrotron emission (dashed line) from a
super--massive black hole $m=10^{9}$ accreting at $\mdot=10^{-3}$
(typical order of magnitude black hole masses and Bondi rate in
ellipticals). The canonical ADAF with $\alpha=0.3$ and $\beta=0.5$ is
unequivocally ruled--out by our radio/sub--mm limits: it greatly
overestimates the total possible contribution at these frequencies.

ADAF models have been applied to a number of accreting black holes
systems: galactic black holes candidates, (e.g NBM, ENM) the centre of
our galaxy (Sgr. A; Narayan et al. 1998) and low luminosity galactic
nuclei (Lasota 1996, Reynolds et al. 1996, Di Matteo \& Fabian 1997)
All of those models seemed to be consistent with the choice of
$\alpha=0.3$ and $\beta=0.5$ (exact equipartition). The canonical ADAF
model (with such a choice of $\alpha$ and $\beta$) has been regarded
as a one parameter model (e.g NBM); the only free parameter (adjusted
to fit the X-ray data) being the mass accretion rate $\mdot$ of the
different systems. This cannot be applied to these low luminosity
elliptical galaxy systems any longer. As we shall show in the course
of this section, the severe discrepancy between the radio and sub--mm
observations with the ADAF model predictions raises serious issues
about the explanation for low--luminosity systems.  Although the
present VLA and SCUBA observations do not rule out the presence of an
ADAF in these elliptical galaxies cores, they do place new (very
tight) constraints on the physical properties of any proposed ADAF,
the plausibility of which remains an object of discussion. Further
studies might be necessary to fully address the different factors
regulating low--luminosity systems.

For all three elliptical galaxies analyzed here we now have a black
hole mass determination from Magorrian et al. (1998), shown in Table
~\ref{t:bh}, and their respective accretion rates estimated from the
Bondi rate in Eqn. 1. By invoking exact equipartition and $\alpha \sim
0.3$, as done in previous work, our systems are virtually
parameter--free; but we are unable to obtain consistency with the
radio data (see dashed lines in Fig.~\ref{4649spec} and
~\ref{4472_4636specs}. Because of the apparent consistency with the
(although looser) optical and X-ray limits we are prompted to
hypothesise that either a lower magnetic field strength might be the
cause of the lack of synchrotron emission or the emission is
free--free absorbed by clumps of cold and dense matter. Also a
magnetic wind or outflow might operate in these systems and remove
most of the mass and angular momentum at large distances so that the
central densities and emissivities are much smaller than in a standard
ADAF.  We will analyze mainly these three possibilities and keep $m$ and
$\mdot$ (apart from Section 4.4) as set from above (the thin solid
line in Fig. ~\ref{4649spec} shows a model for which $\beta\approx
0.5$, $\alpha \approx 0.3$, i.e. the 'standard' ADAF parameters, and
where $\mdot$ is varied as to fit the radio limits.  We find that in
order to be consistent with the radio measurements the accretion rate
has to be very (implausibly) small, $\mdot \approx 10^{-6}$
i.e. $\dot{M} \sim 10^{-5} \Msun {\rm yr}^{-1}$).

Because of the jet-dominated emission in NCG 4472 and the lack of any
high energy component in NGC 4636, we mainly concentrate our
modelling on NGC 4649 which is strongly core dominated and whose
spectrum shows indications of a sharp turnover. As shown in Fig. 
~\ref{4649spec} and ~\ref{4472_4636specs} the computed spectrum has
five major peaks.  From the left these correspond to self--absorbed
thermal synchrotron emission, doubly--Comptonized synchrotron emission,
bremsstrahlung emission and $\gamma$--rays from neutral pion decay.

\subsection{High $\beta$ ADAFs}
\begin{figure*}
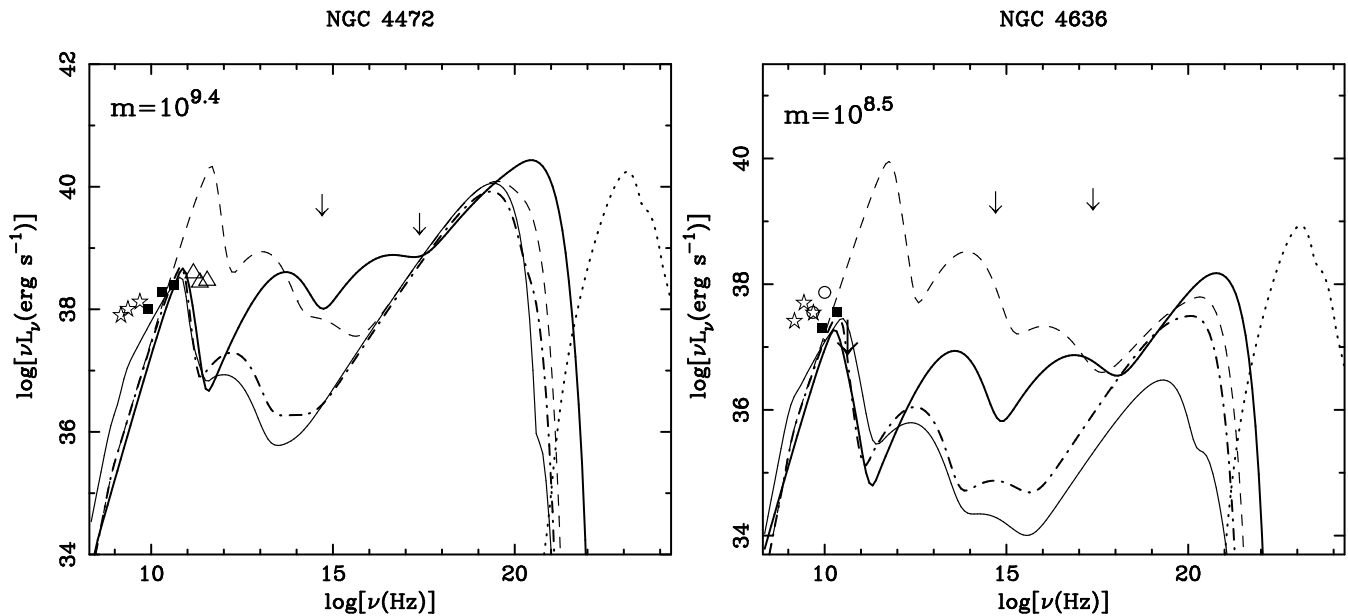

\centerline{
\hbox{
\psfig{figure=fitall_n4472.ps,width=0.5\textwidth,angle=270}
\psfig{figure=fitall_n4636.ps,width=0.5\textwidth,angle=270}
\newline
}}
\caption{
The full band spectra calculated with an ADAF model for NGC 4472 (on
the left) and NGC 4636 (on the right).  Black hole masses, given in Table
~\ref{t:bh}, with the respective Bondi rates and the viscosity parameter,
$\alpha=0.3$, have not been changed in the different models.  Four
models are shown: (i) one for equipartition magnetic field, $\beta=0.5$, 
dashed line, clearly inconsistent with the radio limits (ii) one for
$\beta =0.999$ (NGC 4636) and $\beta =0.99$ for NGC 4472, solid line,
(iii) one for which the radio emission is free--free absorbed
within $r_{\rm in}=30$ (NGC 4636) and $r_{\rm in}=10$ for NGC 4472,
dashed--dot line. Also (iv) the magnetic wind model with $\mdot 
\propto (r/r_{\rm max})^{p}$. The thin solid lines are for $r_{\rm max}=90$ 
and $p=1.2$ (NGC 4472), and for $r_{\rm max}=350$ and $p=0.9$ (NGC
 4636). Distances of $15.8\Mpc$ and $16\Mpc$ are assumed for NGC 4472
 and NGC 4636 respectively. X-ray cooling processes dominate the
 emission.}
\label{4472_4636specs}
\end{figure*}

The solid lines in Fig.~\ref{4649temp_specfig}, ~\ref{4649spec} and
~\ref{4472_4636specs} demonstrate the comparison between an ADAF
model, with very low magnetic field, and the data. As $\beta$ is
increased to $\approxlt 0.99$ ($B\propto (1 - \beta)^{1/2}$, see
Eqn. (2)) the agreement with the radio limits improves significantly.
The decrease in magnetic field strength with respect to the
equipartition value is quite severe. This is required by the fact that
the luminosity (approximated by the Rayleigh--Jeans limit) at the
self--absorbed frequency peak $\nu_{\rm c} (\propto B T_{\rm e}^2$)
scales as
\begin{equation} 
\nu_{\rm c} L_{\nu c} \propto \alpha^{-3/2} (1-\beta)^{3/2} T_{\rm e}^7 m^{1/2} \mdot^{3/2},  
\end{equation} 
i.e. the power output is a very strong
function of $T_{\rm e}$, which increases as $\beta$ increases. With the
inclusion of advection for the electrons, the low $\mdot$,  $\mdot \ll
\mdot_{\rm crit}$, flows are effectively adiabatic and the electron
temperature is basically determined by the balance between the
increase by adiabatic compression and the decrease by radiative
cooling (Nakamura et al. 1997, NBM). As a result of this, for low
magnetic fields ($\beta \approxlt 1$) the temperature increases almost
monotonically (Fig. ~\ref{4649temp_specfig}). For strong fields the rise
of $T_{\rm e }$ is suppressed by very efficient synchrotron and
Compton cooling and the temperature in the inner regions of the flow
is almost constant (see Fig. ~\ref{4649temp_specfig}).

The solid line in Fig.~\ref{4649spec} shows the model for NGC 4649 in
which $B$ is $\sim 1$ per cent of the equipartition magnetic field. The
model is consistent with the high energy VLA and SCUBA upper limit. It
is fairly discrepant with the Fabbiano et al. lower frequency
measurements where the contribution from a jet might become more
relevant (Fig.~\ref{radiomaps}). If all of the emission were to
originate from (or be dominated by) a separate jet component, the
particle energy distribution would need to be very sharply cut off in
order for the emission practically to die out between $5 \times
10^{10}$ Hz and $\sim 10^{11}$ Hz, as implied by our high energy VLA
and SCUBA constraints (see also discussion at the end of Section
4.3). As mentioned before, the large spectral break can be naturally
accounted for by thermal synchrotron radiation in ADAF models. A
similar discussion could also apply to NGC 4636 (Fig.~\ref
{4472_4636specs}), where a relatively sharp spectral turnover seems to be
occurring at $\sim 10^{10}$ Hz. (Although the lack of any higher energy
measurements for this objects make the situation less clear).  Also, an
upper limit ADAF contribution to the much flatter, probably
jet--originating, radio spectrum of NGC 4472 limits $B$ to be within a
few percent of the equipartition value.

Because the system is highly photon starved and $T_{\rm e} \approxgt
10^{10}\K$ the Comptonized spectrum is extremely hard (see solid lines
in Fig. ~\ref{4649spec} and Fig.~\ref{4472_4636specs}) showing clearly the two
Compton scattering orders. In this regime of $\mdot \sim 10^{-3}$,
bremsstrahlung emission usually dominates the X-ray emission and
becomes increasingly important for higher black hole masses.

As shown in Eqn. (4) the peak luminosity is also inversely
proportional to $\alpha$. Higher values for the viscosity parameter
would also decrease the luminosity and therefore allow for lower
values of $\beta$ (still implying a magnetic field strength $B$ 
lower then equipartion).

Although such low magnetic energy densities, as invoked here, do not
pose any real problem to the dynamics of an ADAF (which is supported
by gas pressure) they appear quite implausible for astrophysical
systems. Also, it is hard to reconcile the high viscosity parameter
$\alpha \sim 0.3$ with the magnetic pressure becoming so small.
According to recent studies (e.g. Balbus \& Hawley 1996) $\alpha$
should be linearly correlated with the magnetic field in the disk
(although it not clear at all how $\alpha \sim 0.3$ can be obtained
from magnetic shear instabilities in an ADAF, which also saturates for
magnetic fields in equipartition; the low magnetic field can still act
as a catalyst for turbulence but could hardly be important enough to
make $\alpha$ as high as it is required). The high viscosity could
still be justified if global processes, such as a wind would operate
to transport angular momentum away from the system.
\begin{table*}
\caption{Summary of data for the core of NGC 4472.}
\label{t:4472}
\begin{center}
\begin{tabular}{ccccc}\hline
Frequency & $\nu F_{\nu}$ & reference & notes \\
$\nu$ (Hz) & (10$^{-15}$\,erg\,s$^{-1}$\,cm$^{-2}$) & \vspace{0.3cm} \\
\hline 
$1.5\times 10^9$ & 3.36 & Wrobel (1991)  & VLA \\
$2.3\times 10^9$ & 3.50 &  Wrobel (1991)  & VLA \\
$5.0\times 10^9$ & 4.76 & Wrobel (1991) & VLA\\
$8.4\times 10^9$  & 3.70 & this work & VLA\\
$2.2\times 10^{10}$ & 6.82 & this work & VLA \\
$4.3\times 10^{10}$ & 8.90 & this work &VLA \\
$1.5\times 10^{11}$ & 14.0 & this work & SCUBA \\ 
$2.2\times 10^{11}$ & 9.9 &  this work & SCUBA \\ 
$3.5\times 10^{11}$ & 10.5 &  this work & SCUBA \\ 
$5.45\times 10^{14}$ & $\le$ 180 & Byun et al. (1996) & HST \\
$2.4\times 10^{17}$  & $\le$ 68 & this work & {\it ROSAT} HRI\\
\hline
\end{tabular}
\end{center}
\end{table*}
\begin{table*}
\caption{Summary of data for the core of NGC 4636.}
\label{t:4636}
\begin{center}
\begin{tabular}{ccccc}\hline
Frequency & $\nu F_{\nu}$ & reference & notes \\
$\nu$ (Hz) & (10$^{-15}$\,erg\,s$^{-1}$\,cm$^{-2}$) & \vspace{0.3cm} \\
\hline 
$1.5\times 10^9$ & 0.88 & Wrobel (1991)  & VLA \\
$2.3\times 10^9$ & 1.74 &  Wrobel (1991)  & VLA \\
$4.75\times 10^9$ & 1.23 & Fabbiano  et al. (1987), Wrobel (1991) & Effelsberg, VLA\\
$8.4\times 10^9$  & 0.70 & this work & VLA\\
$1.07\times 10^{10}$ & 2.57 & Fabbiano et al. (1987) & Effelsberg \\
$2.2\times 10^{10}$ & 1.23 & this work & VLA \\
$4.3\times 10^{10}$ & $\le$0.77 & this work &VLA \\ 
$5.45\times 10^{14}$ & $\le$ 68.7 & Byun et al. (1996) & HST \\
$2.4\times 10^{17}$  & $\le$ 78.1 & this work & {\it ROSAT} HRI\\
\hline
\end{tabular}
\end{center}
\end{table*}
~~~
~~~~~~~

\subsection{Free--free absorption of synchrotron radiation by cold material}
A small amount of very dense matter in the central region of an ADAF
have an important effect on the emitted spectrum (Celotti, Fabian \&
Rees 1992). In particular, if a small fraction of the total amount of
matter is in the form of small dense clouds much of the primary
synchrotron radiation emitted by an ADAF would be absorbed.  As
examined by Ferland \& Rees (1988) the radiative equilibrium of very
dense gas irradiated by an intense source is dominated by free--free
absorption which is extremely effective at depleting the low frequency
radiation from the primary source (in this case the ADAF). The
absorbed primary radiation is re--emitted as quasi--blackbody
radiation at the equilibrium temperature.  In other words, the
magnetic field in an ADAF could be as high as in equipartition but the
synchrotron emission suppressed.  As discussed by Rees (1987) the
relatively strong magnetic field in the flow, responsible for the
production of the synchrotron radiation, can also provide the means of
cloud confinement. The theoretically simplest possibility is that the
gas is confined to small regions from which the magnetic field is
excluded. The condition of pressure equilibrium (i.e. cloud thermal
pressure balancing the compressive magnetic stresses) then imposes a
maximum density $n_{\rm c} \sim 10^{15} T_4^{-1} B_2^2\pcmcu$ where a
reference value of $T_4 \approxlt 10^4 \K$ has been assumed for the
temperature of the gas in the clouds and $B_2=10^2 B_2$G is the
magnetic field in the vicinity of the hole from Eqn. (2) with $\beta
\sim 0.5$ and $\alpha=0.3$, $B_2=2(r/3)^{-5/4} (\mdot/10^{-3})^{1/2}
(m/10^9)^{1/2}$ (see also Celotti \& Rees 1998 for more details).
With such density and temperature and absorption extending up to
frequency $10^{12}
\nu_{12}$ Hz the thickness of the absorbing region can be as small as
$h\sim 30 T_4^{1/2}n_{\rm c}^{-2}\nu_{12}^3(1-{\rm e}^{-5\times
10^{-3}\nu_{12}/T_4})\cm$.  A small volume filling factor can still be
consistent with a large covering fraction if the clouds are very
numerous, or if their thickness is much smaller then their length,
i.e. filaments.

In our ADAF model, the synchrotron radiation at different frequencies
is produced at different radii in the flow, the emission at $\sim
10^{12}$ Hz (see dashed lines in Fig.~\ref{4649spec} and
Fig.~\ref{4472_4636specs}) coming from close to the black hole and the
emission from lower frequencies coming from further out. It is
plausible that the free--free absorption can be large enough to
deplete most of the high energy emission within a certain radius. (If
the clouds are sheared as they are accreted inward, their column
density could go down. But because $n_{\rm c}$ increases in the higher
pressure regions free--free opacity would still go up so that 
small--$r$ absorption can be even more dominant).

The dash--dotted lines in Fig.~\ref{4649spec} and
~\ref{4472_4636specs} show a simple representation of such a model
where $\beta =0.5$, $\alpha=0.3$, i.e. the 'canonical' ADAF values,
and where the inner radius at which $f_{\rm c} \sim 1$ has been
constrained from the VLA and SCUBA measurements to be between $r_{\rm
in} \sim 15$ to 30 depending on the particular
source. Fig. ~\ref{4649spec} also shows representative blackbody
emission originating from the cold gas in the clouds (the detailed
spectrum of the reprocessed and cloud emission is beyond the scope
of this paper; it has been calculated in more detail by Kuncic, Celotti
\& Rees 1997).  

Alternatively, within the context of this hypothesis, most of the
emission could be free--free absorbed by cold material in the disk and
what we observe is only the radio emission from the small jets present
in these objects. In this connection, we note that Falcke (1996; also
Becker \& Duschl 1997) has proposed non--thermal models for the radio
emission in Sgr. A$^*$ assuming that most of the electrons have Lorenz
factors of around a few hundred. By requiring that the nonthermal
electrons have monoenergetic electron distribution, such a model would
be able to reproduce the sharp cut--offs implied by our VLA and SCUBA
observations of NGC 4649 and NGC 4636 (the nonthermal synchrotron
usually has a spectral form $L_{\nu} \sim \nu^{-0.7}$; this means that
the optically thin emission would otherwise continue to rise).

We note that any mechanism which removes (or hides) the inner part of
the ADAF emission will suffice to bring the model into agreement. An
ADAF down to $r \sim r_{\rm in}$ and a Bondi flow down to the smallest
radii. Time--variability is also a possibility, but it needs to be very
extreme. This might be related to the large--scale random motions
associated with dissipation in ADAFs (Blackman 1998) and (fairly
implausibly) would need to affect the three objects systematically in
the same direction. More promisingly, as we shall discuss below, a
magnetic wind and/or an outflow could suppress the emission
from the inner regions of the flow.

\subsection{Magnetic wind / outflow}
It has been shown (Blandford \& Payne 1982) that energy, angular
momentum and mass can be removed magnetically from accretion disks, by
field lines that leave the disk and extend to large distances.  Within
the context of a self--similar solution it is possible for such a
magnetically driven wind to carry away most of the mass and angular
momentum at large distances so that the central densities pressures
and emissivities can be smaller that of a standard ADAF (Begelman \&
Blandford in preparation); the flow can then be geometrically thinner
than the standard ADAF. We find (Fig.~\ref{4649spec},
~\ref{4472_4636specs}) that by taking $\mdot\propto r^{p}$ with $p\sim
1$ most of the emission from the inner regions of the flow is
suppressed and the model is brought into agreement with the
observations. The presence of a magnetically driven wind or some kind of
an outflow not only seems plausible in this context, but may also be
necessary in order for the ADAF solution to have a Bernoulli constant
less than zero. As remarked by Narayan \& Yi (1995) in the self--similar
ADAF solution the Bernoulli parameter (defined as the sum of the
kinetic energy, the potential energy and the enthalpy of the accreting
gas) is positive implying that gas on any streamline can spontaneously
escape to infinity. (In contrast to the standard thin disk, in the ADAF
case radiation is suppressed and the excess energy production is
stored as internal energy, making the Bernoulli constant positive).

If most of the mass (not just the angular momentum) can be extracted
 at large distances by magnetic torques, the fraction of mass that
 flows into the very central regions can be quite small. In
 Fig.~\ref{4649spec} and~\ref{4472_4636specs} we show how the radio
 data can be explained by such a flow (a hydromagnetic wind can
 potentially be totally non--radiating; Blandford \& Payne 1982) when
 we adopt a simple model in which the accretion rate decreases as $r$
 decreases. We take $\mdot = \mdot_{\rm Bondi}$ down to a maximum
 radius $r_{\rm max}$ and $\mdot =\mdot_{\rm Bondi} (r/r_{\rm max})^p$
 from $r_{\rm max}$ all the way into the centre, where typical values
 of $p$ and $r_{\rm max}$ are $\sim 1$ and $ \sim $ a few $100$,
 respectively.  Figs.~\ref{4649spec} and ~\ref{4472_4636specs}
 illustrate how a decreasing $\mdot$ greatly affects the high
 frequency synchrotron emission component (and its Comptonization) and
 not so drastically the bremsstrahlung (as much of its contribution
 comes from outer parts of the flow). Note, though, that UV/X-ray
 measurements would be able to discriminate between this model from
 those in the previous Sections (4.2 and 4.3). Also, because of the
 additional dependence of $\mdot$ on $r$ the slope of the
 self--absorbed synchrotron radiation flattens as $p$ increases. In
 the case of NGC 4649 therefore the slope of the observed spectrum
 sets an upper limit to $p$ which cannot much exceed $\sim 1$ (see
 Fig.~\ref{4649temp_specfig}) so that $r_{\rm max} \sim 300$.
 Finally, because at large distances from the disc, the inertia of the
 gas can cause the magnetic field to become increasing toroidal, the
 magnetic stresses could be responsible for converting the centrifugal
 outflow into a more collimated jet structure (observed in the radio,
 see Fig. ~\ref{radiomaps}. The flow might be more similar to an
 ion--supported torus (Rees et al. 1982) which includes outflows in
 funnels. If an outflow does occur (as seems to be the case for NGC
 4472 and NGC 4636) it must be fairly well collimated in order that it
 does not invalidate our accretion estimates. It is possible that an
 outflow makes the accretion rate unsteady.

\section{Summary \& Discussion}
Previous work (Fabian \& Rees 1995, Reynolds et. al. 1996, Di Matteo
\& Fabian 1997, Mahadevan 1997) has suggested that accretion of hot gas in an
elliptical galaxy may create the ideal circumstances for an ADAF to
operate. The present observational studies, aimed at finding further
evidence for of such accretion mode, have shown that new strong
physical constraints need to be introduced to the proposed central
ADAF in such systems.  We have examined new high frequency VLA and
sub--mm SCUBA observations of the three giant, low--luminosity
elliptical galaxies NGC 4649, NGC 4472 and NGC 4636. At these
frequencies the ADAF model predicts that the emission is dominated by
self--absorbed thermal synchrotron radiation. Our new radio limits
show a severe discrepancy with the canonical ADAF predictions and
significantly overestimate the observed flux.  While the present
observations do not rule out the presence of an ADAF in the systems
considered, they do place strong constraints on the model. We have
discussed how the emission could be suppressed because either (1) the
magnetic field in the flow is extremely low (if the viscosity
parameter $\alpha$ can still be high enough) or (2) synchrotron
emission is free--free absorbed by cold material in the accretion
flow. Finally, (3) we discuss how slow non--radiating accretion flows
may drive winds to remove energy, angular momentum and mass so that the
central densities, pressure and emissivities are much smaller than in
a standard ADAF. As most of the accreting gas is lost at large distances,
this scenario becomes particularly appealing for explaining very low
luminosity systems. The present observations raise serious issues
about the explanation for the quiescence of elliptical galaxies nuclei
and possibly other nearby systems now known to posses quiescent
supermassive black holes.

The validity of our findings is strengthened by recent radio
observations of another source, NGC 4258 (Herrnstein et al. 1998),
whose accretion is also modelled via an ADAF (Lasota et al. 1996). The
case of NGC 4258 is particularly illustrative: because the relative
position of the centre of mass of the sub--parsec molecular disk can
be measured to a fraction of a milliarcsecond with VLBI, NGC 4258 has
provided a rare opportunity to circumvent the ambiguity associated
with core--jet emission and to test the radio predictions of the ADAF
spectrum directly. The potential synchrotron emission in that case
also overestimates the maximum possible contribution allowed by the 22
GHz upper limit of Herrnstein et al. The authors show that the
potential ADAF must occur within $r=10^2$. Our two proposed
possibilities for the suppression of synchrotron emission would still
be viable for NGC 4258. In particular, the shearing in the accretion disk
might lead to the disruption of cold dense clouds as the material is
accreted inward. The covering fraction might in this case be high
enough even in the outer regions, to suppress the lower energy
synchrotron radiation. Clearly in the case of NGC 4258 higher
frequency radio measurements (probing the small--$r$ emission and a
possible spectral turnover) could easily test the plausibility of
such an hypothesis.

In conclusion, the compelling observational evidence for ADAFs in
low--luminosity systems, which could have potentially been found
through high radio frequency and sub-mm observations, is still
missing.  Theoretically, the existence of advection-dominated solution
is based on certain critical assumptions, the plausibility of which
still needs to be assessed fully. Although ADAFs have been promisingly
applied to a great variety of low--luminosity systems, and
convincingly to our galactic centre (Mahadevan 1998, NBM), questions
concerning their universality must necessarily be addressed.  Recent
development of ADAF models represents an important step towards the
understanding of low efficiency accretion, but further studies are
necessary to fully address the different factors regulating low--
luminosity systems.  X-ray {\it AXAF} studies would be crucial for
potentially resolving the faint point sources in the nuclei of these
elliptical galaxies and therefore providing clues for understanding
the fuelling of nearby massive black holes and for potentially
discriminating between the different physical constraints on ADAFs we
have discussed in the course of this study.

\section*{Acknowledgements}
We are very grateful to Juri Poutanen for providing us a code to
calculate inverse Compton spectra and to Steve Allen for carrying out
the deprojection analysis. We thank Ramesh Narayan, Eric Blackman and
Rohan Mahadevan for discussions. TDM acknowledges PPARC and Trinity
College, Cambridge, for support. ACF thanks the Royal Society for
support.  RJI is supported by a PPARC Advanced Fellowship. The ASTERIX
software package has been used for the X--ray data analysis.
 The National Radio
Astronomy (NRAO) is a facility of the National Science Foundation,
operated under cooperative agreement by Associated Universities, Inc.

\appendix 
\section{Gamma--ray emission from ADAFs}
The production of $\gamma$--rays from proton proton collisions is a
two step process and has been calculated in detail by Stecker (1971,
and also Stephens \& Badhawar 1981; Dermer 1986ab). The colliding
protons produce a neutral pion $\pi^0$, which then decays into two
$\gamma$-rays. Within the context of ADAF models pion emission can
become relevant and it has been calculated by Mahadevan et al. (1997).
Here, we consider a power law distribution of proton energy described
by a spectral index $p$ (as expected if energy if viscous energy is
dissipated directly into the protons; for pion production for a
thermal distribution see Di Matteo 1998).
The pion spectrum is then given by:
\begin{eqnarray}
F(E_{\pi})&=&
\frac{I(p,\alpha,\beta)m\mdot}{m_{\pi}}\\ \nonumber 
&&\int_{1}^{\infty}\gamma^{-p}\
(1-\gamma)^{-1/2}\frac{d\sigma_{\rm
mb}(\gamma_{\pi},\gamma)}{d\gamma_{\pi}}d\gamma \\ \nonumber
&&\hspace{2cm}{\rm photons \,\,s^{-1} GeV^{-1}}
\end{eqnarray}
where $I(p,\alpha,\beta)m\mdot$ is given in Mahadevan et al. 1997 and
it includes the integral over radius of the energy distribution
function and the density of protons. $\gamma$ is the Lorentz factor,
$m_{\pi} = 0.135 {\rm Gev}/c^2$ and $d\sigma_{\rm
mb}(\gamma_{\pi},\gamma)/d\gamma_{\pi}$ is the differential
cross--section determined by the isobar model (Stecker 1971) and the
scaling model (Stephens and Badhawar 1981) used in the respective
ranges.  

To obtain the $\gamma$--ray emission, we calculate the
integrals above and use the relation (Stecker 1971),
\begin{equation}
F(E_{\gamma})= 2\int_{E_{\rm \pi min}}^{\infty}\frac{F(E_{\pi})}{\sqrt{E_{\pi}^2 - m_{\pi}^2}} ,
\end{equation}
in photons $\ps$ GeV$^{-1}$, and where $E_{\pi{\rm min}}$ is the
minimum pion energy required to produce a $\gamma$--ray with energy
$E_{\gamma}$,
\begin{equation}
E_{\pi{\rm min}} = E_{\gamma} + \frac{m_{\pi}^2}{4E_{\gamma}^2}.
\end{equation}

\end{document}